\documentclass[twocolumn,showpacs,showkeys,preprintnumbers,amsmath,amssymb]{revtex4}

\usepackage{graphicx}
\usepackage{dcolumn}
\usepackage{bm}
\usepackage{epsfig}

\begin{document}
\preprint{IST/CFP 5.2005}

\title[]{EHD ponderomotive forces and aerodynamic flow control using plasma actuators}

\author{Mario J. Pinheiro}
\affiliation{Department of Physics and Center for Plasma Physics,
Instituto Superior T\'{e}cnico, Av. Rovisco Pais,
 1049-001 Lisboa, Portugal}
\email{mpinheiro@ist.utl.pt}

\homepage{http://alfa.ist.utl.pt/~pinheiro}

\thanks{This work was partially financed by Funda\c{c}\~{a}o Calouste Gulbenkian and the Rectorate of the Technical
University of Lisbon. I would like to thank Prof. John Reece Roth
for many helpful conversations and for the opportunity to stay as
an invited research scholar in the Plasma Laboratory at the
University of Tennessee, Knoxville.}





\date{\today}

\begin{abstract}
We present a self-consistent two-dimensional fluid model of the
temporal and spatial development of the One Atmosphere Uniform Glow
Discharge Plasma (OAUGDP$^{\circledR}$). Continuity equations for
electrically charged species N$_2^+$, N$_4^+$, O$_2^+$, O$_2^-$ and
electrons are solved coupled to the Poisson equation, subject to
appropriate boundary conditions. It was used an algorithm proposed
by Patankar. The transport parameters and rate coefficients for
electrons at atmospheric pressure are obtained by solving the
homogeneous Boltzmann equation for electrons under the hydrodynamic
assumption. Operational variables are obtained as a function of
time: electric current; surface charge accumulated on the dielectric
surface; the memory voltage and the gas voltage controlling the
discharge. It is also obtained the spatial distribution of the
electric field, the populations of charges species, the resulting
ponderomotive forces, and the gas speed.
\end{abstract}

\bibliographystyle{apsrev}

\maketitle


\section{Introduction}

The development of the One Atmosphere Uniform Glow Discharge Plasma
(OAUGDP$^{\circledR}$) has made it possible to generate purely
electrohydrodynamic (EHD) ponderomotive (body) forces
~\cite{Roth1,Roth2,Roth3,Roth4}. Such forces are generated without a
magnetic field and with small intensity currents crossing the
plasma. In fact, only RF displacement currents produce the body
forces that accelerate the plasma. Two methods were devised for flow
acceleration~\cite{Roth95,Roth01}: 1) Peristaltic flow acceleration
and 2) Paraelectric flow acceleration. Only the last method is
analyzed in this work. Paraelectric flow acceleration is the
electrostatic analog of paramagnetism: a plasma is accelerated
toward increasing electric field gradients, while dragging the
neutral gas with it. Applications span from propulsion and control
systems in aeronautics, to killing any kind of bacterium and virus
(see Ref.~\cite{Roth1}).

The role of plasma in aerodynamic research has been increasing,
since it constitutes a significant energy multiplier modifying the
local sound speed and thus leading to modifications of the flow and
pressure distribution around the
vehicle~\cite{Bletzinger,Soldati,Shyy}. Plasma actuators have been
shown to act on the airflow properties at velocities below 50 m/s
\cite{Pons}.

In default of a complete model of a OAUGDP$^{\circledR}$ reactor,
Chen ~\cite{Chen} built a specific electrical circuit model for a
parallel-plate and coplanar reactor, modeling it as a
voltage-controlled current source that is switched on when the
applied voltage across the gap exceeds the breakdown voltage.

Although there is still lacking a detailed characterization of such
plasma actuators, with only boundary layer velocity profiles
measured using a Pitot tube located 1-2 mm above the flat panel
~\cite{Roth04} being available, we present in this paper a
self-consistent two-dimensional modeling of temporal and spatial
development of the OAUGDP$^{\circledR}$ in an "airlike" gas.

\section{Numerical model}

\subsection{Assumptions of the model}

We intend to describe here the glow discharge regime, with emphasis
on flow control applications of the plasma. Gadri~\cite{Gadri} has
shown that an atmospheric glow discharge is characterized by the
same phenomenology as low-pressure dc glow discharges.

No detailed plasma chemistry with neutral heavy species is presently
available; only the kinetics involving electrically charged species
supposedly playing a determinant role at atmospheric pressure:
N$_2^+$, N$_4^+$, O$_2^+$, O$_2^-$, and electrons, is addressed. The
electronic rate coefficients and transport parameters are obtained
by solving the homogeneous electron Boltzmann equation under the
hydrodynamic regime assumption~\cite{Ferreira}. When obtaining the
charged species populations as well the electric field controlling
their dynamics, the following electrohydrodynamics (EHD) effects are
studied in this work:
\begin{itemize}
    \item ponderomotive forces acting on the plasma horizontally and perpendicularly to the energized electrode;
    \item gas velocity.
\end{itemize}

\begin{figure}
  \includegraphics[width=3.5 in, height=3.5 in]{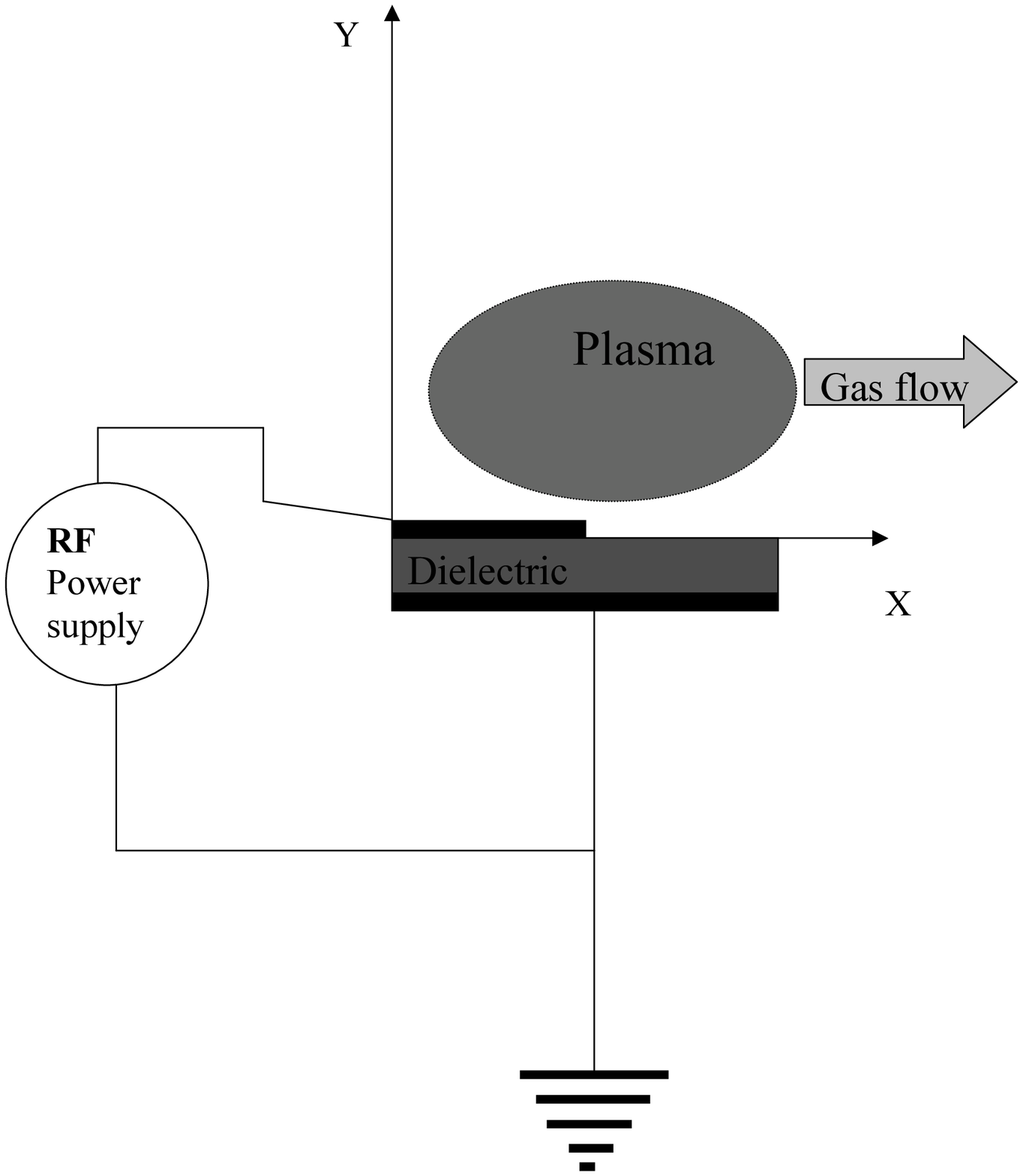}\\
  \caption{Schematic representation of electrode geometry of an energized OAUGDP plasma
  panel.}\label{Fig0}
\end{figure}

The simulation domain is a 2-DIM Cartesian geometry (see
Fig.~\ref{Fig0}) with total length along the X-axis $L_X=0.4$ cm
and height $L_Y=0.4$ cm; the width of the dielectric surface along
the X-axis is 0.3 cm in Case Study I and 0.1 cm in Case Study II.
The dielectric relative permittivity, supposed to be a ceramic
material, is assumed to be $\varepsilon_r=100$; the dielectric
thickness is in all cases set 0.065 cm. The capacity of the
reactor is determined through the conventional formula
$C_{ds}=\varepsilon_0 \varepsilon_r S/d$. The electrodes thickness
is supposed to be negligible.

\subsection{Transport parameters and rate coefficients}

The working gas is a "airlike" mixture of a fixed fraction of
nitrogen ($\delta_{N_2}=[N_2]/N=.78$) and oxygen
($\delta_{O_2}=[O_2]/N=0.22$), as is normally present at sea level
at $p=1$ atm.

The electron homogeneous Boltzmann equation~\cite{Ferreira} is
solved with the 2-term expansion in spherical harmonics for a
mixture of $N_2-22 \%$ O$_2$. The gas temperature is assumed
constant both spatially and in time, $T_g=300$ K, and as well the
vibrational temperature of nitrogen $T_v (N_2)=2000$ K and oxygen
$T_v(O_2)=2000$ K. The set of cross sections of excitation by
electron impact was taken from~\cite{Siglo}.

At atmospheric pressure the local equilibrium assumption holds:
Transport coefficients ($\nu_{ion}^{N_2}$, $\nu_{ion}^{O_2}$,
$\mu_e$, $\mu_p$, $D_e$, $D_p$) depend on space and time
$(\textbf{r},t)$ only through the local value of the electric
field $\mathbf{E}(\mathbf{r},t)$. This is the so called
hydrodynamic regime.

Ion diffusion and mobility coefficients were taken from
\cite{Sigmond}, $\mu_{O_2^-} .N= 6.85 \times 10^{21}$ V$^{-1}$
m$^{-1}$ s$^{-1}$ (on the range of E/N with interest here),
$\mu_{O_2^+} N = 6.91 \times 10^{21}$ V$^{-1}$ m$^{-1}$ s$^{-1}$,
and $\mu_{N_2^+} N = 5.37 \times 10^{21}$ V$^{-1}$ m$^{-1}$
s$^{-1}$.

The reactions included in the present kinetic model are listed in
Table ~\ref{tab:table2}. It is assumed that all volume ionization
is due to electron-impact ionization from the ground state and the
kinetic set consists basically in ionization, attachment and
recombination processes. The kinetics of excited states and heavy
neutral species is not considered.

To obtain a faster numerical solution of the present hydrodynamic
problem it is assumed that the gas flow does not alter the plasma
characteristics and is much smaller than the charged particle
drift velocity. This assumption allows a simplified description of
the flow. For more concise notation, we put $n_{p2}\equiv
[N_4^+]$; $n_{p1}\equiv [N_2^+]$; $n_p \equiv [O_2^+]$; $n_n
\equiv [O_2^-]$, and $n_e \equiv [e]$.

The balance equations for N$_4^+$ (at atmospheric pressure the
nitrogen ion predominant is N$_4^+$) is:
\begin{equation}\label{Eq1}
\frac{\partial n_{p2}}{\partial t} + \mathbf{\nabla} \cdot (n_{p2}
\mathbf{v}_{p2}) = \delta_{N_2}^2 N^2 n_{p1} K_{ic1} - \beta N
n_{p2}- K_{r2} n_{p2} n_e.
\end{equation}
The balance equation for N$_2^+$ is:
\begin{widetext}
\begin{equation}\label{Eq2}
\frac{\partial n_{p1}}{\partial t} + \mathbf{\nabla} \cdot (n_{p1}
\mathbf{v}_{p2}) =  n_e \nu_{ion}^{N_2} + K_{ic2} [N_2] n_{p2} -
\beta_{ii} n_n n_{p1} - \beta n_e n_{p1} - K_{ic1} [N_2]^2 n_{p1}.
\end{equation}
\end{widetext}
The oxygen ion considered is O$_2^+$ and its resultant balance
equation is given by
\begin{equation}\label{Eq3}
\frac{\partial n_{p}}{\partial t} + \mathbf{\nabla} \cdot (n_{p}
\mathbf{v}_{p}) =  n_e \nu_{ion}^{O_2} - \beta_{ii} n_n n_p -
\beta n_e n_p.
\end{equation}
As oxygen is an attachment gas, the negative ion O$_2^-$ was
introduced and its balance equation was written as:
\begin{equation}\label{Eq4}
\frac{\partial n_{n}}{\partial t} + \mathbf{\nabla} \cdot (n_{n}
\mathbf{v}_{n}) = \nu_{att}^{O_2}n_e - \beta_{ii} n_{p1} n_n - K_d
n_p n_n.
\end{equation}
Finally, the balance equation for electrons can be written in the
form:
\begin{widetext}
\begin{equation}\label{Eq5}
\frac{\partial n_{e}}{\partial t} + \mathbf{\nabla} \cdot (n_{e}
\mathbf{v}_{e}) = n_e (\nu_{ion}^{N_2} + \nu_{ion}^{O_2} -
\nu_{att}^{O_2}) - \beta n_e (n_p + n_{p1}) + K_d [O_2]n_n -
K_{r2} n_{p1} n_e.
\end{equation}
\end{widetext}
To close the above system of equations we use the drift-diffusion
approximation for the charged particle mean velocities appearing
in the continuity equations:
\begin{equation}\label{Eq6}
n_i \mathbf{v}_i = n_i \mu_i \mathbf{E} - \nabla (n_i D_i),
\end{equation}
where $\mu_i$ and $D_i$ represent the charged particle mobility
and the respective diffusion coefficient. The applied voltage has
a sinusoidal wave form
\begin{equation}\label{Eq7}
V(t) = V_{dc} + V_0 \sin (\omega t),
\end{equation}
where $V_{dc}$ is the dc bias voltage (although here we fixed to
ground, $V_{dc}=0$) and $\omega$ is the applied angular frequency.
$V_0$ is the maximum amplitude with the root mean square voltage
in this case of study $V_{rms}=5$ kV and the applied frequency
$f=5$ kHz.

The total current (convective plus displacement current) was
determined using the following equation given by Sato and Murray
\cite{Sato}
\begin{widetext}
\begin{equation}\label{Eq8}
I_d(t)=\frac{e}{V} \int_V \left( n_p \mathbf{w}_p - n_e
\mathbf{w}_e - n_n\mathbf{w}_n - D_p \frac{\partial n_p}{\partial
z} + D_e \frac{\partial n_e}{\partial z} + D_n \frac{\partial
n_n}{\partial z} \right) \cdot \mathbf{E}_L d v +
\frac{\epsilon_0}{V} \int_V \left( \frac{\partial
\mathbf{E}_L}{\partial t}\cdot \mathbf{E}_L \right) dv,
\end{equation}
\end{widetext}
where $\int_V dv$ is the volume occupied by the discharge,
$\mathbf{E}_L$ is the space-charge free component of the electric
field. The last integral when applied to our geometry gives the
displacement current component
\begin{equation}\label{Eq9}
I_{disp} (t) = \frac{\varepsilon_0}{d^2} \frac{\partial
V}{\partial t} \int_V dv.
\end{equation}

Auger electrons are assumed to be produced by impact of positive
ions on the cathode with an efficiency $\gamma=5 \times 10^{-2}$,
so that the flux density of secondary electrons out of the cathode
is given by
\begin{equation}\label{Eq10}
\mathbf{j}_{se} (t) = \gamma \mathbf{j}_p (t),
\end{equation}
with $\mathbf{j}_p$ denoting the flux density of positive ions. In
fact, this mechanism is of fundamental importance on the working
of the OAUGDP$^{\circledR}$.

Due to the accumulation of electric charges over the dielectric
surface, a kind of "memory voltage" is developed, whose expression
is given by:
\begin{equation}\label{Eq11}
V_m(t)=\frac{1}{C_{ds}} \int_{t_0}^t I_d (t') d t' + V_m(t_0).
\end{equation}
Here, $C_{ds}$ is the equivalent capacitance of the discharge.

The space-charge electric field was obtained by solving the
Poisson equation
\begin{equation}\label{Eq12}
\Delta V = -\frac{e}{\epsilon_0} (n_p - n_e - n_n).
\end{equation}
The boundary conditions are the following:
\begin{itemize}
    \item electrode (Dirichlet boundary condition): $V(x,y=0,t)=V-V_m$;
    \item dielectric (Neumann boundary condition): $E_n=(\mathbf{E} \cdot
    \mathbf{n})=\frac{\sigma}{2 \epsilon_0}$.
\end{itemize}
The flux of electric charges impinging on the dielectric surface
builds up a surface charge density $\sigma$ which was calculated
by balancing the flux to the dielectric
\begin{equation}\label{Eq13}
\frac{\partial \sigma}{\partial t}=e
(|\Gamma_{p,n}|-|\Gamma_{e,n}|).
\end{equation}
Here, $\mathbf{\Gamma}_{p,n}$ and $\mathbf{\Gamma}_{e,n}$
represent the normal component of the flux of positive and
negative ions and electrons to the dielectric surface.
Furthermore, it is assumed that ions and electrons recombine
instantaneously on the perfectly absorbing surface.

The entire set of equations are solved together, at each time
step, self-consistently.

\begin{table*}
\caption{\label{tab:table2} List of reactions taken into account in
our model. Rate coefficients were taken from Ref.~\cite{Kossyi92}.}
\begin{ruledtabular}
\begin{tabular}{ccc}
 kind of reaction &  Process & Rate coefficient\\
\hline
Ionization & $e + N_2 \to 2e + N_2^+$ &  $\nu_{ion}^{N_2}$ \footnotemark[1]\\
Ionization & $e + O_2 \to 2e + O_2^+$ &  $\nu_{ion}^{O_2}$ \footnotemark[1]\\
3-body electron attachment & $e + O_2 + O_2 \to O_2^- + O_2$ & $K_{a1}=1.4 \times 10^{-29} (\frac{300}{T_e})\exp(-600/T_g) K_1(T_g,T_e)$  (cm$^6/$s) \footnotemark[2] \\
3-body electron attachment & $e + O_2 + N_2 \to O_2^- + N_2$ & $K_{a2}=1.07 \times 10^{-31} (\frac{300}{T_e})^2 K_2(T_g,T_e)$ (cm$^6/$s) \footnotemark[3]\\
Collisional detachment & $O_2^- + O_2 \to  e + 2 O_2$   & $K_d=2.7 \times 10^{-10} \sqrt{\frac{T_g}{300}} \exp(-5590/T_g)$ (cm$^3/$s)  \\
e-ion dissociative recombination & $ N_2^+ + e \to 2N$   &  $\beta=2.8 \times 10^{-7} \sqrt{\frac{300}{T_g}}$ (cm$^3/$s) \\
e-ion dissociative recombination & $O_2^+ + e \to 2O$    & $\beta=2.8 \times 10^{-7} \sqrt{\frac{300}{T_g}}$ (cm$^3/$s) \\
2-body ion-ion recombination & $O_2^- + N_2^+ \to O_2 + N_2$  & $\beta_{ii}=2 \times 10^{-7} \sqrt{\frac{300}{T_g}}[1+10^{-19} N (\frac{300}{T_g})^2]$ (cm$^3/$s) \\
Ion-conversion & $N_2^+ + N_2 + N_2 \to N_4^+ + e$ & $K_{ic1}=5 \times 10^{-29}$ (cm$^6/$s) \\
Recombination & $N_4^+ + e \to 2 N_2$ & $K_{r2}=2.3 \times 10^{-6}/(Te/300)^{0.56}$ \\
Ion-conversion & $N_4^+ + N_2 \to N_2^+ + 2N_2$ &  $K_{ic2}=2.1\times 10^{-16} \exp(T_g/121)$ (cm$^3/$s) \\
\end{tabular}
\end{ruledtabular}
\footnotetext[1]{Data obtained by solving the quasi-stationary,
homogeneous electron Boltzmann equation. See Ref.~\cite{Ferreira}
for details.}

\footnotetext[2]{With $K_1=\exp(700(T_e-T_g)/(T_eT_g))$}
\footnotetext[3]{With $K_2=\exp(-70/T_g)\exp(1500(T_e-T_g)/(T_e
T_g))$}
\end{table*}

\section{Method of resolution of fluid equations}

The particle's governing equations are of convection-diffusion type.
They are solved using a method proposed by Patankar~\cite{Patankar}
(see also Ref.~\cite{Pinhao}). According to this method, let
$L(\phi, d \phi,d^2 \phi,...)=S$ be a homogeneous differential
equation in $\phi$, with a source term $S$. Then the procedure of
Patankar consists in replacing $L(\phi, d \phi,d^2 \phi,...)=S$ by
$a_P \phi_P = \sum_k a_k \phi_k + b$, where P is the central point
of the mesh.

The chosen time step is limited by the value of the dielectric
relaxation time. For the present calculations the total number of
computational meshes used is (100x100). This fair condition allows
calculating an entire cycle with an Intel Pentium 4 (2.66 GHz) in
a reasonable CPU time of about 30 hours per cycle, limiting to a
reasonable value the relative error $|\Delta w_e/w_e|$, with $w_e$
designating the electron drift velocity. Stationarity was attained
typically after 4-5 cycles. Equations~\ref{Eq1}-~\ref{Eq13} are
integrated successively in time supposing the electric field is
constant during each time step, obtaining a new value of the
electric field after the end of each time step. The method used to
integrate the continuity equations and Poisson equation was
assured to be numerically stable, constraining the time step width
to the well known Courant-Levy-Friedrich stability criterion.

\section{Results}

The simulations were done for a two-dimensional flat staggered
geometry, as sketched in Fig.~\ref{Fig0}. This is essentially a
'surface discharge' arrangement with asymmetric electrodes. It is
assumed that the plasma is homogeneous along the OZ axis.

\subsection{Electrical characteristics}

\begin{figure}
  \includegraphics[width=3.5 in, height=3 in]{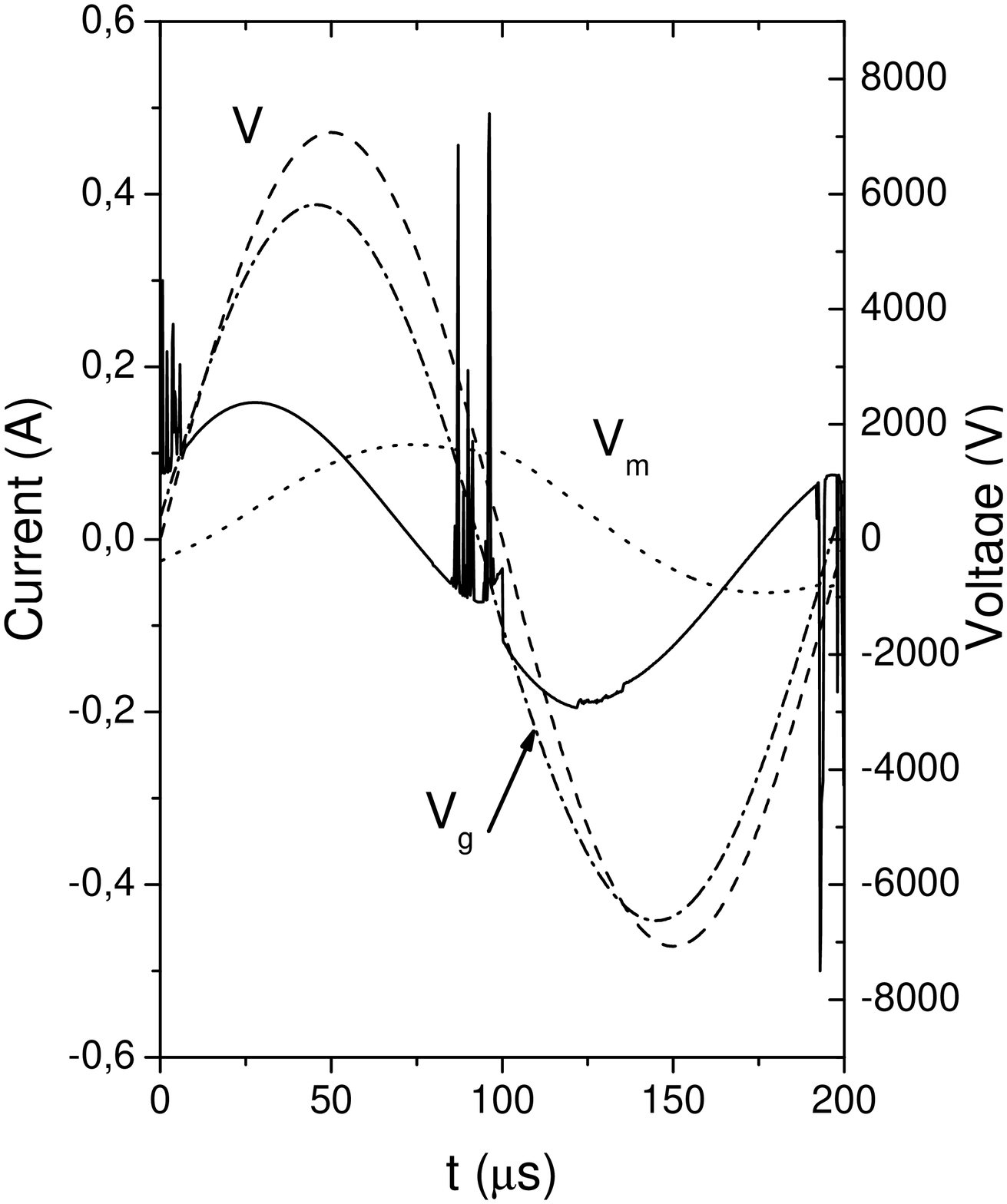}\\
  \caption{Electric current, applied voltage, gas voltage and memory voltage as a function of time.
  Conditions: Case I. Solid curve: current; dot curve: $V_m$; dashed-dot curve: $V_g$; dashed curve: $V$.}\label{fig1}
\end{figure}

In Fig.~\ref{fig1} it is given the evolution along a period of
time of the calculated electric current, applied voltage, gas
voltage and memory voltage. The OAUGDP$^{\circledR}$, and as well
generically a DBD, occurs in configurations characterized by a
dielectric layer between conducting electrodes. At about $740$
Volts, electron avalanches develop, replenishing the volume above
the surface with charged particles. Hence, the charged particles
are flowing to the dielectric (see Eq.~\ref{Eq13}) start
accumulating on the surface, and build-up an electric field that
prevents the occurrence of a high current, and quenches the
discharge development at an early stage.

\subsection{Electrical field and potential}

\begin{figure}
  \includegraphics[width=3 in, height=4 in]{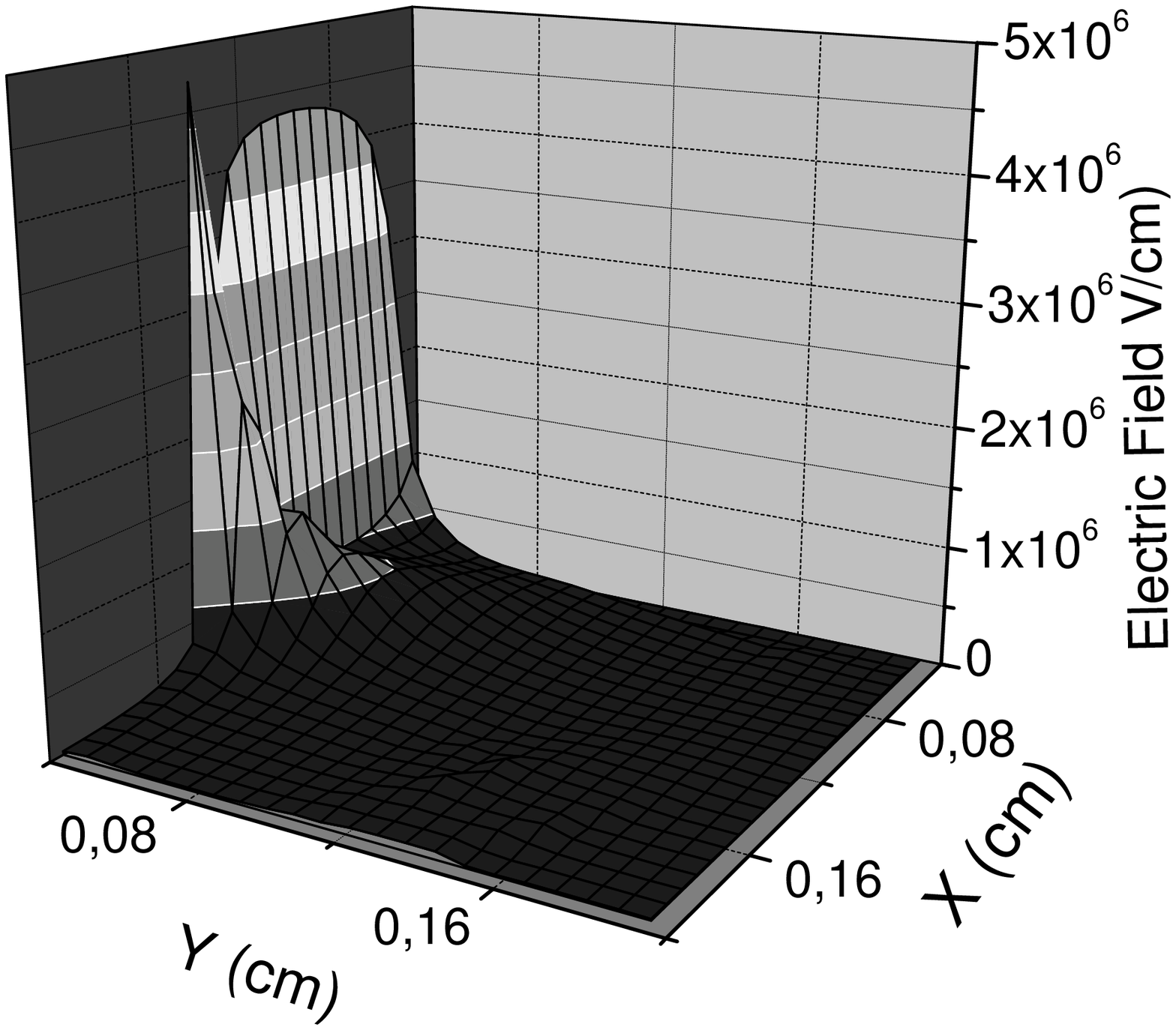}\\
  \caption{Electric field along OX and OY near the energized electrode at time $t=1.9 \times 10^{-5}$ s at first
  half-cycle. Conditions: Case study I, with $V=5$ kV, $f=5$ kHz.}\label{fig9}
\end{figure}

Fig.~\ref{fig9} shows the electric field during the first
half-cycle at the instant of time $t=1.9 \times 10^{-5}$ s. The
energized electrode is the anode and the electric field follows
Aston's law, its magnitude remaining on the order of $10^5$ V$/$cm
at a height of $8 \times 10^{-5}$ m above the electrode and
attaining lower magnitude above the dielectric surface, typically
on the order of $10^3$ V$/$cm. The electric field magnitude is
strongest in region around the inner edges of the energized
electrode and dielectric surface (which is playing during this
half-cycle the role of a pseudo-cathode).

During the avalanche development a strong ion sheath appears. In
fact, as the avalanche develops an ion sheath expands along the
dielectric surface until it reaches the boundary. With the ion
sheath travels an electric field wave, with some similarities with a
solitary wave. The speed of its propagation in the conditions of
Fig.~\ref{fig9} is about 150 m$/$s. See Refs.~\cite{Shyy,Boeuf1} for
very elucidating explanation of this phenomena.

\subsection{Paraelectric gas flow control}

\begin{figure}
  \includegraphics[width=3.5 in, height=3 in]{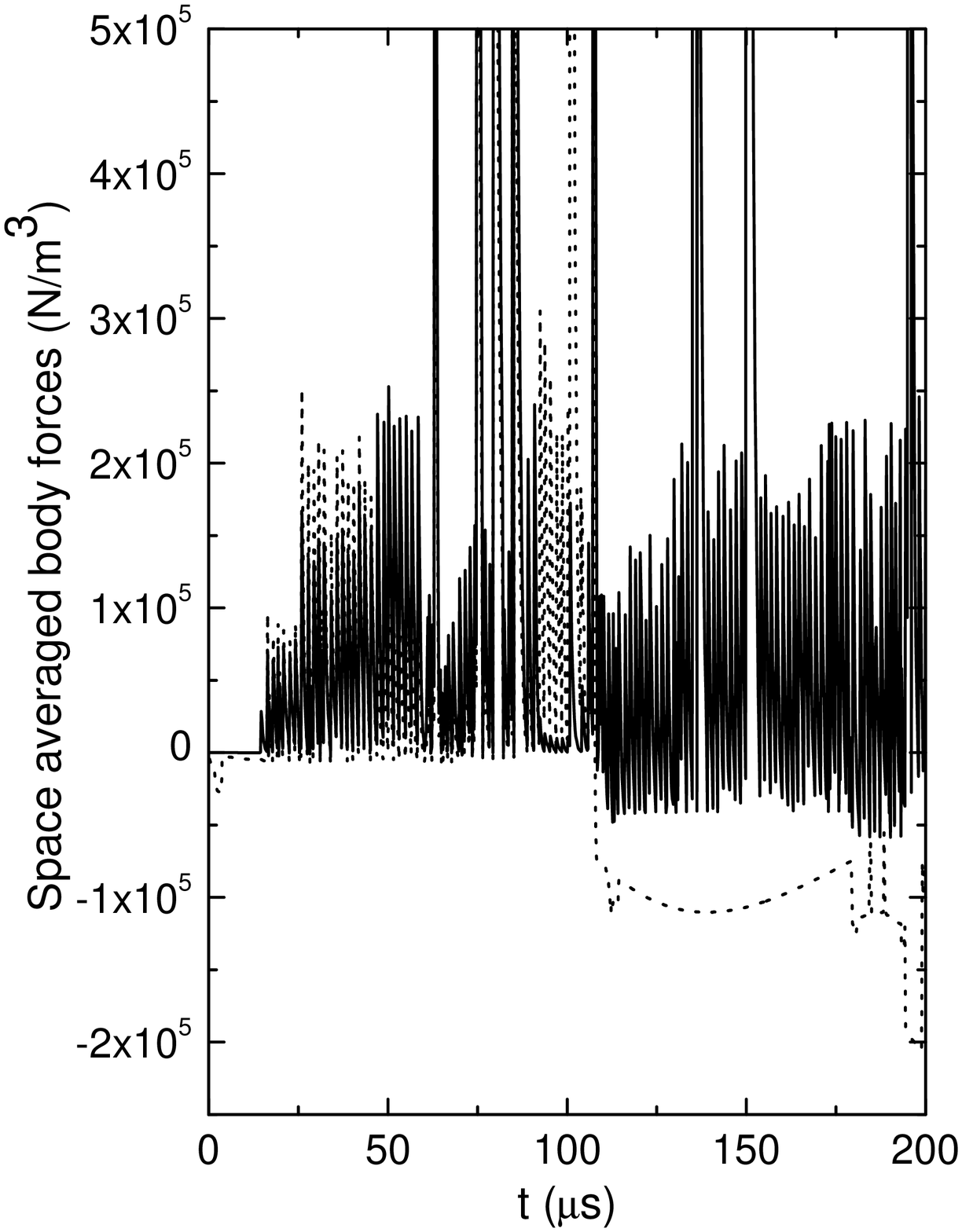}\\
  \caption{Calculated space averaged ponderomotive forces per unit volume as a function of time. Solid curve: $F_x$; dot curve:
  $F_y$. Case study I}\label{fig4}
\end{figure}

\begin{figure}
  \includegraphics[width=3.5 in, height=3 in]{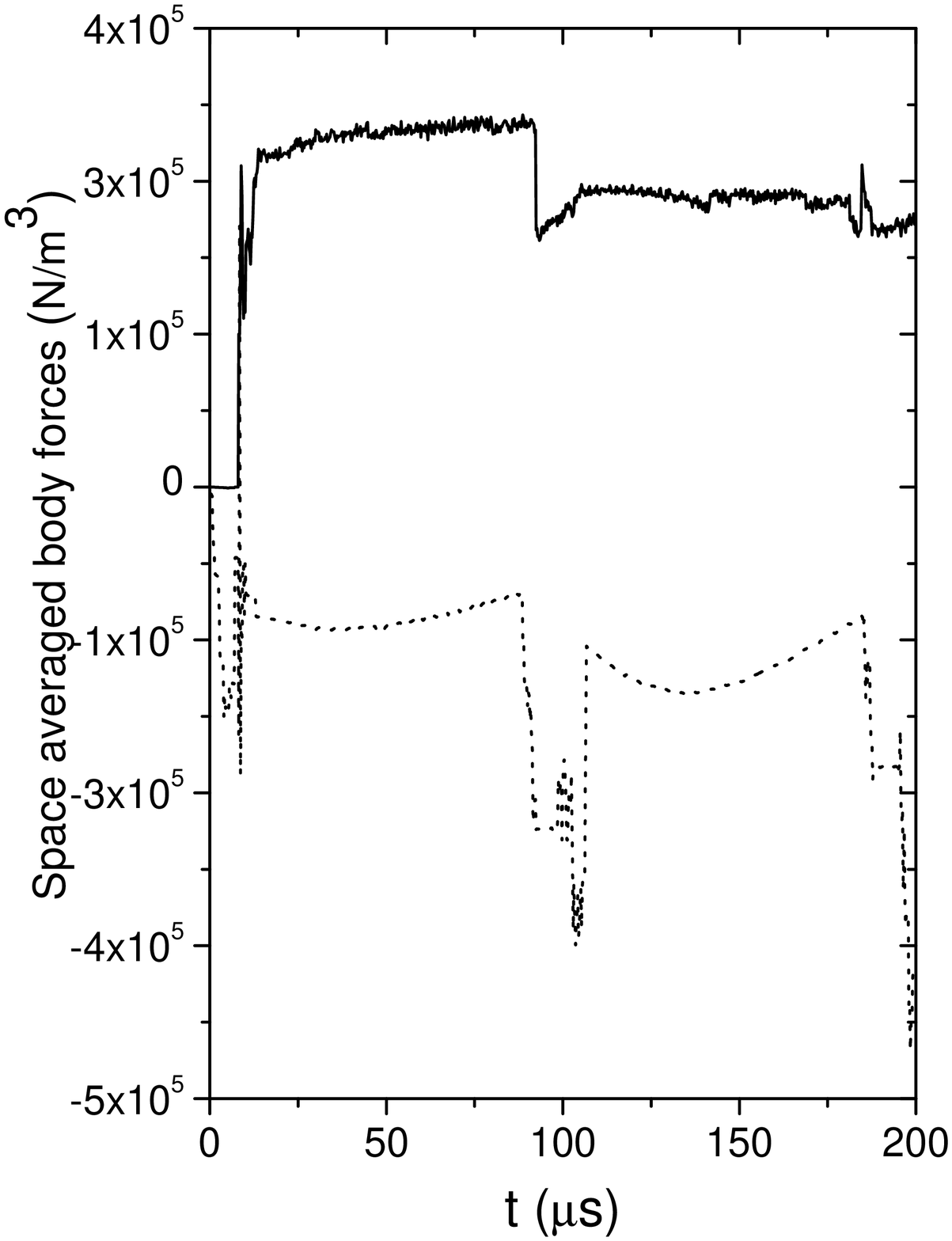}\\
  \caption{Calculated space averaged ponderomotive forces per unit volume as a function of time. Solid curve: $F_x$; dot curve:
  $F_y$. Case study II}\label{fig5}
\end{figure}

The theory of paraelectric gas flow control was developed by
Roth~\cite{Roth1}. The electrostatic ponderomotive force
$\mathbf{F_E}$ (units N$/$m$^3$) acting on a plasma with a net
charge density $\rho$ (units C$/$m$^3$) is given by the
electrostatic ponderomotive force and can be expressed under the
form
\begin{equation}\label{}
\mathbf{F_E}=\frac{1}{2}\varepsilon_0 \nabla E^2.
\end{equation}
In order to verify whether electrostriction effects could be
playing any significant role, it was also calculated the
electrostriction ponderomotive force
\begin{equation}\label{}
\mathbf{F_{es}} = -\frac{1}{2} \varepsilon_0 E^2 \nabla
\varepsilon_r.
\end{equation}
Here, $\varepsilon_r=1-\frac{\omega_p^2}{\nu_{en}^2 + \omega^2}$ is
the relative permittivity of the plasma, $\nu_{en}$ is the
electron-neutral momentum transfer frequency and $\omega_p$ is the
plasma frequency. We found that this force term is negligible,
contributing at maximum with 1 $\%$ to the total ponderomotive
force. Subsequently, the ponderomotive forces were averaged over the
area of calculation. Comparing the calculated space averaged
ponderomotive forces per unit volume shown in
Figs.~\ref{fig4}-~\ref{fig5} it is seen that when the electrode
width increases they become one order of magnitude higher. On
average, during the second half-cycle the ponderomotive force
magnitude decreases. This happens when the voltage polarity is
reversed and the energized electrode play the role of cathode. This
is due to a reduction of the potential gradient on the edge of the
expanding plasma (see also Ref.~\cite{Boeuf1}). Calculations of EHD
ponderomotive force have shown that its maximum intensity is
attained during electron avalanches, with typical values on the
order of $5 \times 10^9$ N$/$m$^3$. $\overline{F}_x$ points along OX
(propelling direction), while $\overline{F}_y$ points downwards
(boundary layer control).

\subsection{Gas speed}

\begin{figure}
  \includegraphics[width=3.5 in, height=3 in]{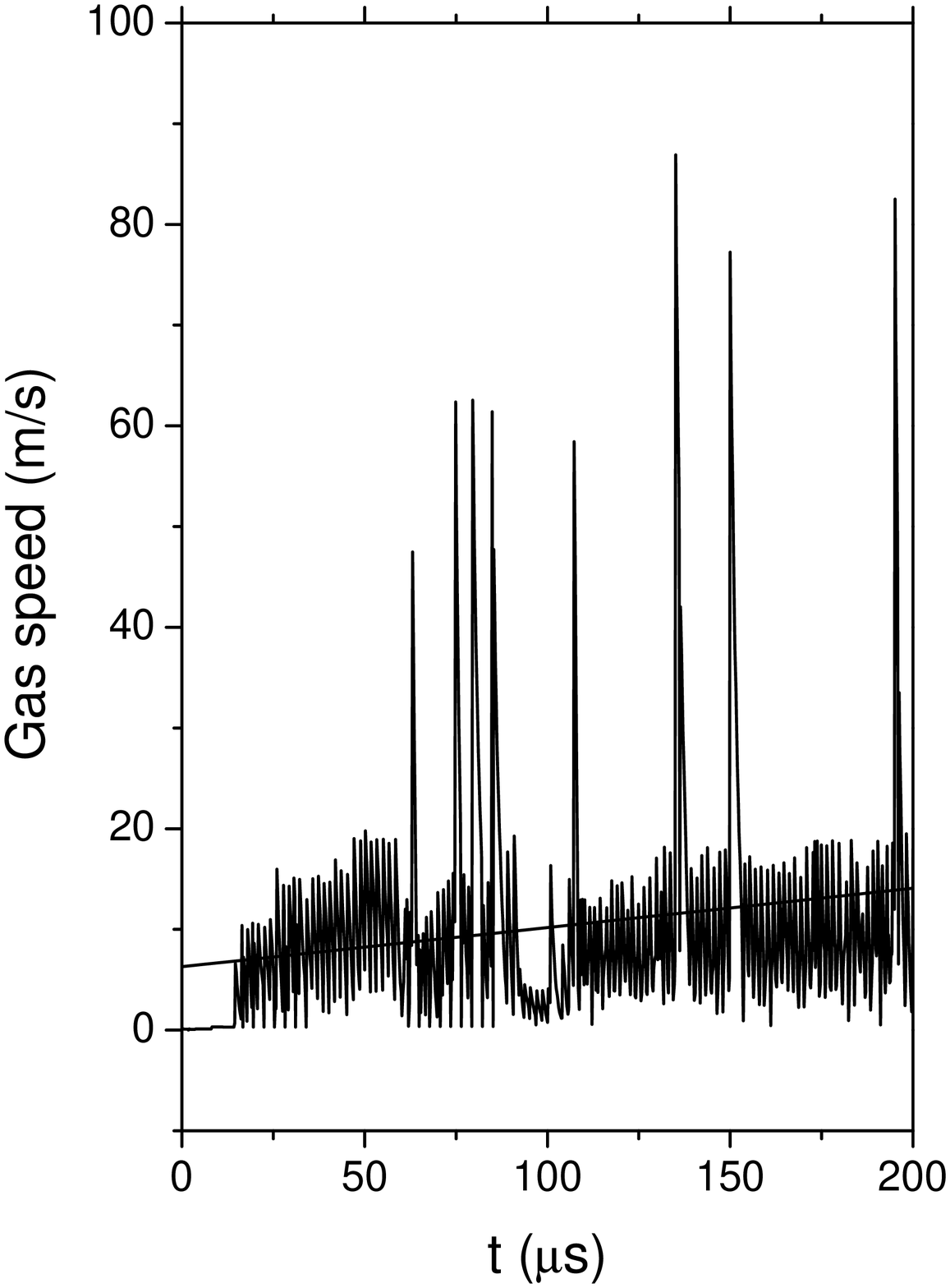}\\
  \caption{Space averaged gas speed as a function of time. The straight solid line is a linear fit showing
  the increase of gas speed with time. Case I}\label{fig11}
\end{figure}

Using Bernoulli law (see Ref.~\cite{Roth1}) it can be obtained the
induced neutral gas speed
\begin{equation}\label{gs2}
v_0 = E \sqrt{\frac{\varepsilon_0 }{\rho}} = \sqrt{\frac{2}{\rho}
\overline{F}_x L_X}.
\end{equation}
Here, $\overline{F}_x$ is the calculated space average ponderomotive
forces per unit volume, and $\rho=1.293$ Kg$/$m$^3$.
Fig.~\ref{fig11} shows the gas speed along the entire cycle in Case
I. The average value of the gas speed is around 15 m$/$s while the
experimental value, measured with a Pitot tube 1-2 mm above the
surface, is 5 m$/$s as high as is for nearly the same operational
conditions~\cite{Roth04}.

\begin{figure}
  \includegraphics[width=3.5 in, height=3 in]{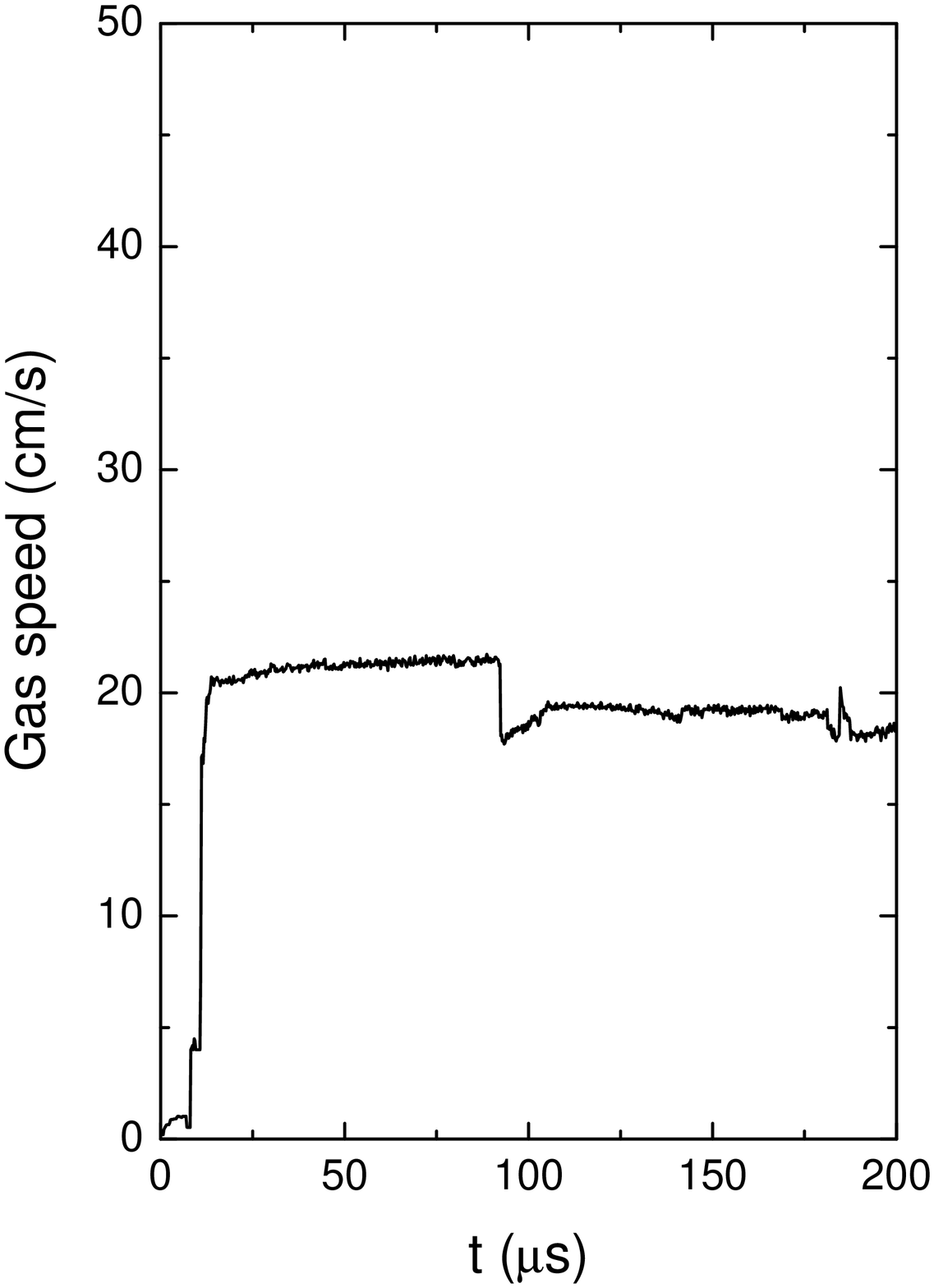}\\
  \caption{Space averaged gas speed as a function of time. Case II}\label{fig12}
\end{figure}

As can be seen in Fig.~\ref{fig12} the gas speed increases to about
20 m$/$s when the dielectric surface decreases. It is clear the
slight decrease of the gas speed during the cathode cycle. This is
related to the decrease of ponderomotive forces, as discussed above.
As we assumed charged particles are totally absorbed on the
dielectric surface, the swarm of ions propagating along the
dielectric surface are progressively depleted, dwindle with time.
However, it is worth to mentioning (see also Ref.~\cite{Khudik})
that in certain conditions the inverse phenomena can happen, a
bigger dielectric width feeding up the ion swarm with newborn ions
and thus inducing an increase of the gas speed. How long its width
can be increased is a matter of further study.

\section{Conclusion}

A 2-DIM self-consistent kinetic model has been implemented to
describe the electrical and kinetic properties of the
OAUGDP$^{\circledR}$. It was confirmed that the electric field
follows the Aston's law above the energized electrode. EHD
ponderomotive forces on the order of $5 \times 10^9$ N$/$m$^3$ can
be generated locally during the electron avalanches, their
intensity decreasing afterwards to values well below on the order
of $10^4 \div 10^5$ N$/$m$^3$. On the cathode side the EHD
ponderomotive forces can decrease $1.5 \div 2$ orders of
magnitude, due probably to a smaller important potential gradient.
The ponderomotive forces (and as well the gas speed) tend to
increase whenever the energized electrode width augments
relatively to the dielectric width.

This code will help to design an advanced propulsion system,
achieving flow control in boundary layers and over airfoils by EHD
means, with numerous advantages over conventional systems.


\end{document}